\documentclass[prd,twocolumn,showpacs,preprintnumbers,aps,nofootinbib]{revtex4-1}

\usepackage{graphicx}
\usepackage{dcolumn}
\usepackage{bm}
\usepackage{amsmath,amssymb,amsfonts}
\usepackage{latexsym}

\usepackage{color}

\begin{document}


\title{\boldmath
Collider Prospects for Muon $g-2$ in General Two Higgs Doublet Model
}

\author{Wei-Shu Hou$^1$}
\author{Rishabh Jain$^1$}
\author{Chung Kao$^2$}
\author{Girish Kumar$^1$}
\author{Tanmoy Modak$^3$}

\affiliation{
$^1$Department of Physics, National Taiwan University, Taipei 10617, Taiwan}
\affiliation{
$^2$Homer L. Dodge Department of Physics and Astronomy,
University of Oklahoma, Norman, OK 73019, USA}
\affiliation{
$^3$Institut f\"ur Theoretische Physik, Universit\"at Heidelberg, 69120 Heidelberg, Germany}
\bigskip


\begin{abstract} 
Recent progress on muon $g-2$ measurement prompts one to take it 
even more seriously. In the general two Higgs doublet model that 
allows extra Yukawa couplings, we take a simplified approach of 
single enhanced coupling. We fix the charged lepton flavor violating 
coupling, $\rho_{\tau\mu} = \rho_{\mu\tau}$, via the one-loop 
mechanism, for illustrative masses of the heavy scalar $H$ 
and pseudoscalar $A$, where we assume $m_A = m_{H^+}$. 
Since extra top Yukawa couplings are plausibly the largest,
we turn on $\rho_{tt}$ and find that LHC search for
$gg \to H,\,A \to \tau\mu$ gives more stringent bound than from 
$\tau\to \mu\gamma$ with two-loop mechanism. 
Turning on a second extra top Yukawa coupling, $\rho_{tc}$, 
can loosen the bound on $\rho_{tt}$, but LHC constraints can again 
be more stringent than from $B \to D\mu\nu$ vs $De\nu$ universality.
This means that evidence for $H,\,A \to \tau\mu$ may yet emerge with 
full LHC Run~2 data, while direct search for
$\tau^\pm\mu^\mp bW^+$ or $t\bar cbW^+$ (plus conjugate) may also bear fruit.
\end{abstract}

\maketitle


\section{Introduction}

After extended, meticulous efforts, the Fermilab Muon $g -2$ experiment
announced recently their first measurement~\cite{Abi:2021gix},
$a_\mu({\rm FNAL}) = 116 592 040(54) \times 10^{-11}$\,(0.46\,ppm).
This confirms the previous result~\cite{Bennett:2006fi} at Brookhaven 
National Laboratory, combining to give~\cite{Abi:2021gix}
\begin{align}
a_\mu({\rm Exp}) = 116 592 061(41) \times 10^{-11}\,(0.35\,{\rm ppm}).
\label{g-2_2021}
\end{align}
Comparing this with the ``consensus'' theory prediction~\cite{Aoyama:2020ynm} 
for the Standard Model (SM), namely $a_\mu({\rm SM}) = 
116 591 810(43) \times 10^{-11}$\,(0.37\,ppm), the difference 
\begin{align}
a_\mu({\rm Exp}) - a_\mu({\rm SM}) =  (251 \pm 59) \times 10^{-11},
\label{da_mu}
\end{align}
is at 4.2$\sigma$. Eq.~(\ref{g-2_2021}), however, is consistent with 
a new lattice result~\cite{Borsanyi:2020mff} based on staggered fermions. 
Thus, the issue of the true SM value remains.
We shall take Eq.~(\ref{da_mu}) as is and 
seek $1\sigma$ solution with New Physics.

The persistence of the ``muon $g-2$ anomaly'' means there is 
a truly vast theory literature, hence we refer to a very recent 
comprehensive account~\cite{Athron:2021iuf} for more complete references. 
Ref.~\cite{Athron:2021iuf} stresses the need for chiral enhancement 
in solving the muon $g-2$ anomaly, whereas the two Higgs doublet model (2HDM) 
is ``the only possibility without introducing new vector bosons or leptoquarks".
We shall follow the 2HDM that is ``not flavor-aligned''~\cite{Athron:2021iuf},
which possesses extra Yukawa couplings such as charged lepton flavor violating (CLFV)
$\tau$-$\mu$ couplings, namely Ref.~\cite{Iguro:2019sly}
 (which descends from 
 Refs.~\cite{Assamagan:2002kf,Davidson:2010xv,Omura:2015nja,Omura:2015xcg,Iguro:2017ysu,Crivellin:2019dun}). 
We expand on the impact at the Large Hadron Collider (LHC)
by considering extra top Yukawa couplings~\cite{Hou:2020chc}.

The well-known 2HDM~
 Model I and II invoke 
a $Z_2$ symmetry to implement the Natural Flavor Conservation (NFC) 
condition of Glashow and Weinberg~\cite{Glashow:1976nt}, 
i.e. just one Yukawa matrix per quark charge
 (and for charged leptons as well).
But this is ``special'', if not {\it ad hoc}, so in the {\it general} 2HDM (g2HDM)
one drops the $Z_2$ symmetry and let {\it Nature} reveal her flavor design. 
First called Model III~\cite{Hou:1991un}, and following 
the footsteps of the Cheng-Sher ansatz~\cite{Cheng:1987rs}, indeed
the emergent fermion mass mixing hierarchies can be exploited to ease  
the worries~\cite{Glashow:1976nt} of flavor changing neutral couplings (FCNC): 
extra Yukawa matrices should trickle off when going off-diagonal.
The recent emergent alignment phenomenon, that the observed 
$h$ boson at 125 GeV resembles very closely~\cite{Khachatryan:2016vau} 
the SM Higgs boson, brought in a flavor-independent surprise: 
alignment suppresses~\cite{Hou:2017hiw} FCNC involving the $h$ boson.
{\it Nature}'s designs for flavor seem intricate.

\begin{figure}[b]
\center
\includegraphics[width=0.22 \textwidth]{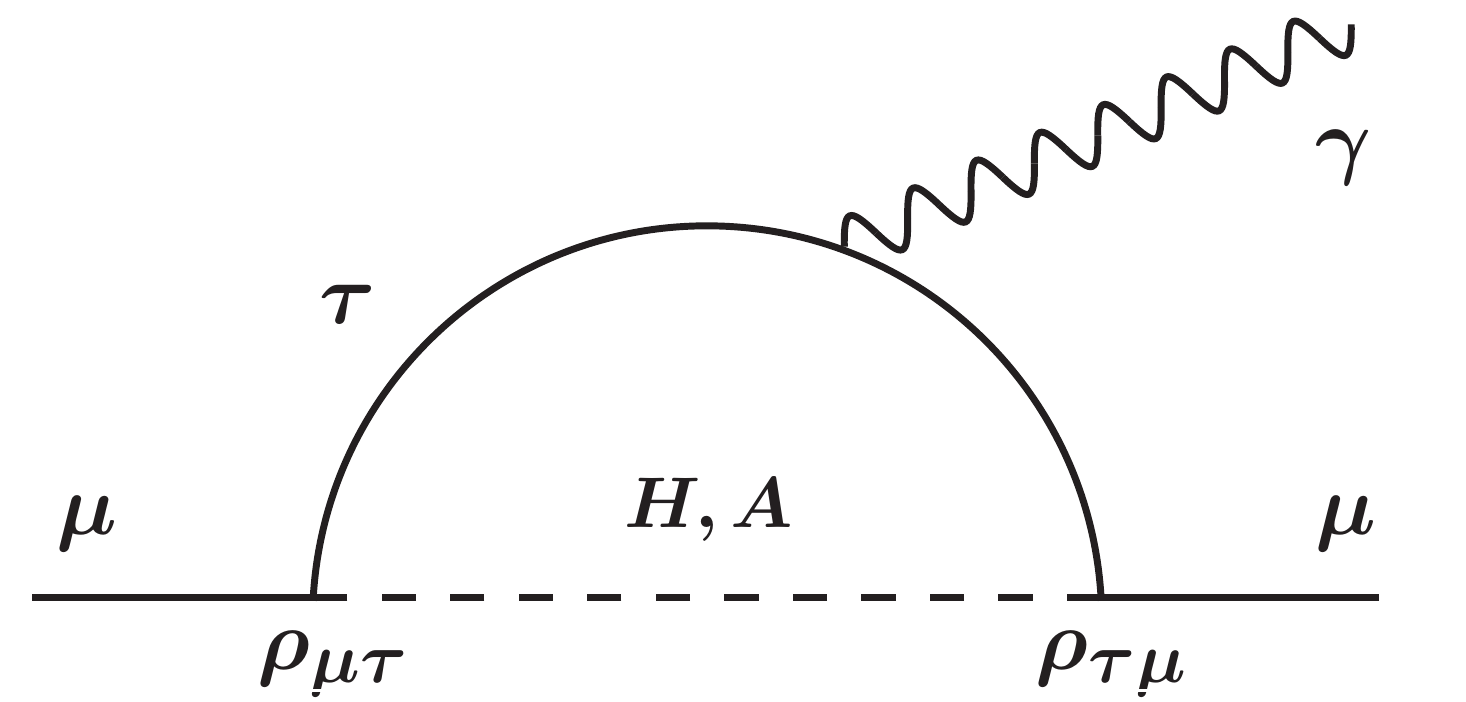}
\hskip0.1cm
\includegraphics[width=0.22 \textwidth]{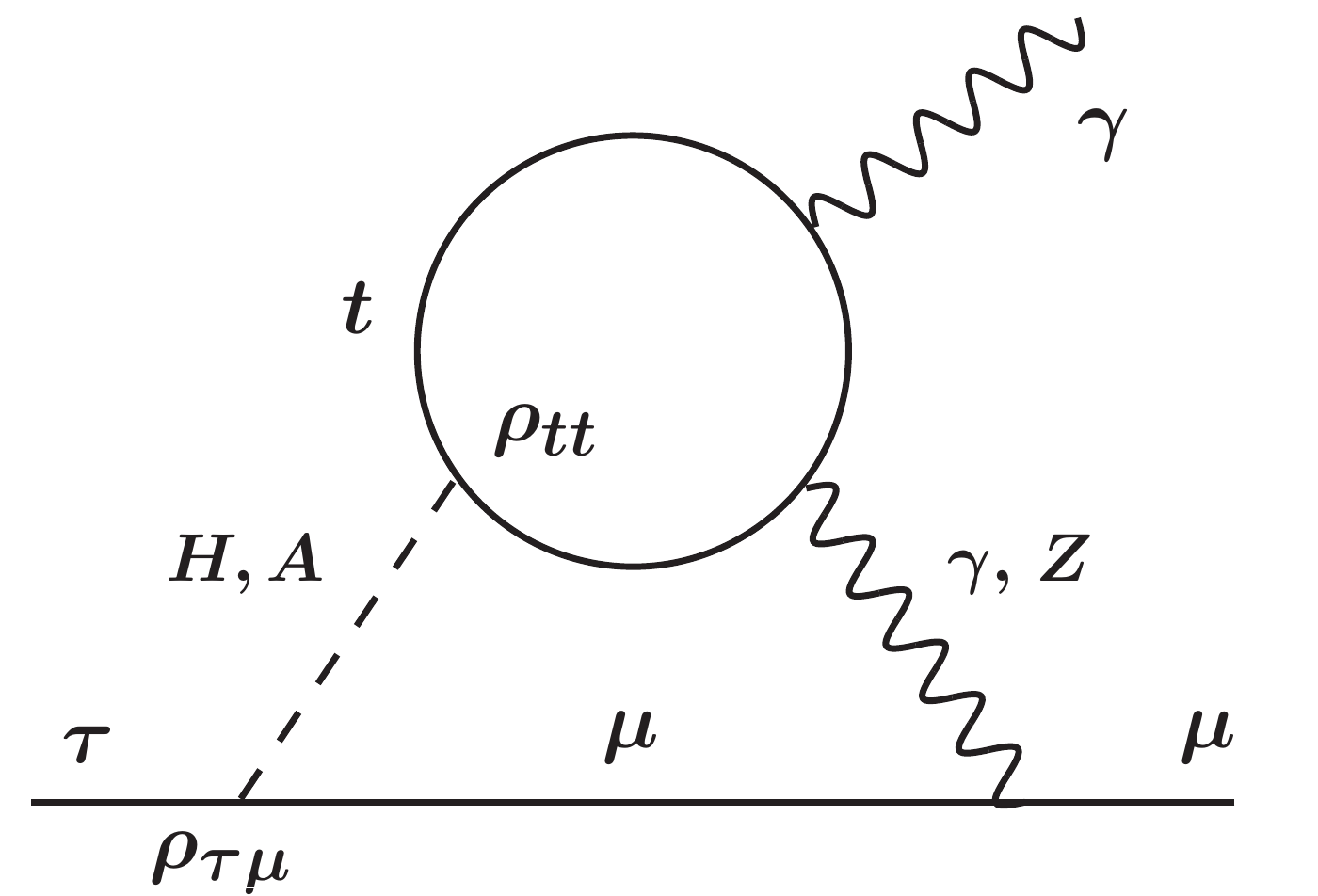}
\caption{
One-loop mechanism for muon $g-2$, and two-loop diagram for $\tau\to \mu\gamma$.}
\label{feyndiag}
\end{figure}

The alignment control of FCNC is illustrated by $h \to \tau\mu$ search.
The CMS experiment found initially~\cite{Khachatryan:2015kon} 
an intriguing 2$\sigma$ hint, 
which subsequently disappeared~\cite{PDG}.
The full Run~2 data at 13 TeV gives~\cite{Sirunyan:2021ovv},
\begin{align}
{\cal B}(h \to  \tau\mu) < 0.15 \%. \quad\quad ({\rm CMS2021})
\label{h-taumuCMS}
\end{align}
But since the FCNC $\rho_{\tau\mu}$ arises from the heavy exotic
doublet $\Phi'$ ($\langle \Phi' \rangle = 0$)
rather than the mass-giving doublet $\Phi$ (sole source of vacuum
expectation value),
the $h\tau\mu$ couples as $\rho_{\tau\mu}c_\gamma$,
where $c_\gamma \equiv \cos\gamma$ is the $h$--$H$ mixing angle
between the two $CP$-even scalar bosons.
Thus, alignment, that $c_\gamma$ or $h$--$H$ mixing is small, can account 
for Eq.~(\ref{h-taumuCMS}) without requiring $\rho_{\tau\mu}$ to be small, 
which is analogous~\cite{Chen:2013qta} to another FCNC process, $t \to ch$
 {(with coupling $\rho_{tc}c_\gamma$)}.
This is the starting point for a one-loop mechanism (see Fig.~\ref{feyndiag}(left))
to account for the muon $g-2$ anomaly~\cite{Omura:2015nja,Omura:2015xcg},
originally stimulated by the CMS hint~\cite{Khachatryan:2015kon} for $h \to \tau\mu$. 
Refs.~\cite{Iguro:2019sly,Iguro:2017ysu} followed up on further LHC implications.

Invoking the one-loop mechanism of Fig.~\ref{feyndiag}(left) to account for 
muon $g-2$ for $m_{A,\,H}$ at the weak scale implies rather large $\rho_{\tau\mu}$~\cite{Iguro:2019sly,Assamagan:2002kf,Davidson:2010xv,Omura:2015nja,Omura:2015xcg,Iguro:2017ysu}, 
at several tens times $\lambda_\tau \cong 0.01$, the tau Yukawa coupling in SM.
Eq.~(\ref{h-taumuCMS}) then demands $|c_\gamma| \ll 1$, 
i.e. near the alignment limit.
On one hand this calls for a symmetry, which we do not get into. 
On the other hand, one should turn to
$H,\, A \to \tau\mu$ (and $H^+ \to \tau^+\nu_{(\mu)},\, \mu^+\nu_{(\tau)}$)
search~\cite{Hou:2019grj}, as it is not hampered by small $c_\gamma$ 
but at full strength of $s_\gamma\;(\equiv \sin\gamma) \to -1$.
This needs finite $\rho_{tt}$ for gluon-gluon fusion production,
so let us articulate our approach.

With one Higgs doublet of the SM already fully affirmed, adding 
a second doublet is the most conservative and simple extension.
But, while simple, without a $Z_2$ symmetry to enforce NFC, 
the g2HDM possesses {\it many} new parameters. We therefore take 
a {\it simplified} approach of one large extra Yukawa coupling at a time, especially 
if it is greatly enhanced compared with analogous SM couplings.
{By analogy with the known top Yukawa coupling, $\lambda_t \cong 1$, 
however, it seems plausible that the extra top Yukawa coupling 
$\rho_{tt}$ is the strongest.}
{Taking $\rho_{\tau\tau} \lesssim {\cal O}(\lambda_\tau)$ 
to keep the one-loop effect small,}
the large $\rho_{\tau\mu}$ needed for muon $g-2$ 
can induce $\tau\to\mu\gamma$ with finite $\rho_{tt}$~\cite{Davidson:2010xv}
through the two-loop diagram of Fig.~\ref{feyndiag}(right), which 
places a bound on $\rho_{tt}\rho_{\tau\mu}$, where the Belle experiment 
has recently updated~\cite{Abdesselam:2021cpu} with full data.
This constrains $\rho_{tt}$ to be considerably smaller 
than the $\rho_{\tau\mu}$ needed for muon $g-2$.
We will show that a recent search~\cite{Sirunyan:2019shc} for 
$gg \to H \to \tau\mu$ by CMS with 36 fb$^{-1}$ data at 13 TeV,
when interpreted in g2HDM, would place bounds on $\rho_{tt}$ 
that are {more stringent than from Belle}. 

%

{To enlarge the allowed range for $\rho_{tt}$, we turn on 
a second extra top Yukawa coupling, $\rho_{tc}$,}
{which can dilute~\cite{Iguro:2019sly} the branching ratio ${\cal B}(H,\,A \to \tau\mu)$,
thereby extend the allowed range for $\rho_{tt}$.}
However, the product of $\rho_{tc}\rho_{\tau\mu}$ can
induce $B \to D\mu\nu_\tau$ through $H^+$ exchange and affect 
the measured $B \to D\mu\nu$ rate (as the $\nu_\tau$ flavor cannot be detected), 
thereby break universality~\cite{Iguro:2019sly} with $B\to De\nu$.
Depending on the ${H^+}$ mass, 
$\rho_{tc}$ comparable in strength to $\rho_{\tau\mu}$ can be allowed.
We will argue that searching for
$\tau^\pm\mu^\mp bW^+$ or $t\bar cbW^+$ at the LHC should be of interest.

The purpose of this paper is therefore threefold. First, g2HDM 
{\it can}~\cite{Iguro:2019sly,Assamagan:2002kf,Davidson:2010xv,Omura:2015nja,Omura:2015xcg,Iguro:2017ysu}
account for muon $g-2$ anomaly, which is not new.
Second, possible astounding signatures may emerge soon, not just at Belle~II,
but also {\it at the LHC}, if large $\rho_{\tau\mu}$ is behind the muon $g-2$ anomaly. 
Our two LHC contact points above imply that $gg \to H,\,A \to \tau\mu$ 
may suddenly emerge, perhaps even with full Run~2 data,
or detailed work may discover novel signatures
such as $\tau^\pm\mu^\mp bW^+$ or $t\bar cbW^+$ at the LHC.
Third, this illustrates how limited our knowledge of g2HDM really is.

In Sec.~II we discuss the one-loop mechanism and 
{find the bound on $\rho_{\tau\mu}$, then discuss
flavor constraints on $\rho_{tt}$ and $\rho_{tc}$ that follow, 
as well as} various flavor concerns;
in Sec.~III we compare $\tau\to\mu\gamma$ constraint on $\rho_{tt}$
with direct search for $gg \to H,\, A \to \tau\mu$, and
$B \to D\ell\nu$ universality ($\ell = e,\,\mu$) constraint {on $\rho_{tc}$} 
with $\tau^\pm\mu^\mp bW^+$ or $t\bar cbW^+$ search at the LHC;
after some discussion in Sec.~IV, we offer our summary. 

\section{\boldmath
One-loop mechanism for muon $g-2$}

The Yukawa couplings in g2HDM are~\cite{Davidson:2005cw,Hou:2019mve}
\begin{align}
& - \bar{\nu}_i\rho^\ell_{ij} R \, \ell_j H^+
    - \bar{u}_i\left[(V\rho^d)_{ij} R - (\rho^{u\dagger}V)_{ij} L\right]d_j H^+ \notag\\
& - \frac{1}{\sqrt{2}} \sum_{f = \ell,}^{u, d} \bar f_{i}\Big[
 \Big.\big(\lambda^f_i \delta_{ij} c_\gamma + \rho^f_{ij} s_\gamma\big)H
 - i\,{\rm sgn}(Q_f) \rho^f_{ij} A \Big. \notag\\
 &\quad\quad\quad\quad\quad
  - \big(\lambda^f_i \delta_{ij} s_\gamma - \rho^f_{ij} c_\gamma\big) h \Big]  R\, f_{j}
 +{h.c.},
\label{eq:Yuk}
\end{align}
where $i$, $j$ are summed over generations, 
$L, R = (1\mp\gamma_5)/2$ are projection operators, $V$ is the CKM matrix, 
with lepton matrix taken as unity due to vanishing neutrino masses.
One can therefore read off the $\rho_{\mu\tau}$, $\rho_{\tau\mu}$ 
and $\rho_{tt}$ couplings indicated in Fig.~\ref{feyndiag}.

\begin{figure}[t]
\center
\includegraphics[width=0.35 \textwidth]{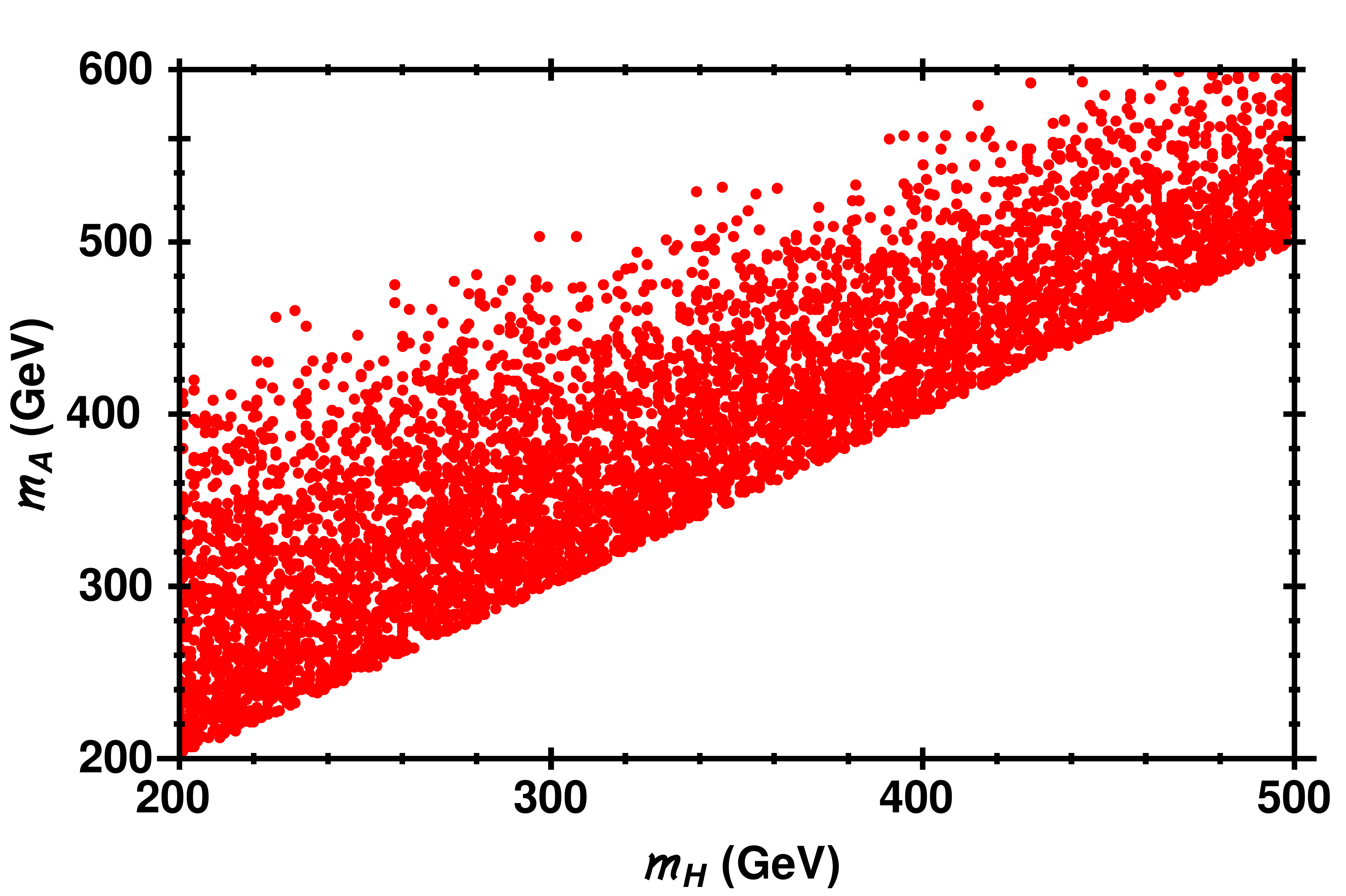}
\caption{
Parameter space in $m_H$--$m_A$ that satisfy
perturbativity, unitarity, positivity and oblique electroweak constraints.}
\label{mHmA_scan}
\end{figure}

We do not write down the Higgs potential $V(\Phi,\, \Phi')$ 
(except assuming it is $CP$ conserving), 
as it can be found in many papers traced to Ref.~\cite{Davidson:2005cw}.
We shall take $m_H = 300$ GeV as benchmark to keep $\rho_{\tau\mu}$ ``reasonable'', 
and illustrate with $m_A = m_{H^+} = 340,\, 420,\, 500$~GeV.
That is, we assume custodial symmetry
 (namely $\eta_4 = \eta_5$, in the notation of Ref.~\cite{Hou:2017hiw})
to reduce tension with oblique parameter constraints.
We follow, for example, Ref.~\cite{Hou:2021xiq} and perform a parameter scan 
utilizing 2HDMC~\cite{Eriksson:2010zzb} to demonstrate that there is 
parameter space 
that satisfy perturbativity, unitarity and positivity 
as well as precision electroweak constraints,
as shown in Fig.~\ref{mHmA_scan}.
The scan is performed with $m_H = [200,\, 500]$~GeV, 
$m_A = m_{H^+} = [200,\, 600]$~GeV,  $
\eta_2 = [0,\, 5]$ (as required by positivity), 
$\eta_3 = [-5,\, 5]$ and $\eta_7 = [-5,\, 5]$,
and setting the alignment limit $\cos\gamma = 0$ with $m_A > m_H$.

As we will soon see, to provide 1$\sigma$ solution to the muon $g-2$ anomaly 
for benchmark $m_H$, $m_A$ values,  $\rho_{\tau\mu} = \rho_{\mu\tau}$
 (which we take to simplify)
need to be $\sim 20$ times the strength of $\lambda_\tau \simeq 0.01$.
Eq.~(\ref{h-taumuCMS}) then implies 
\begin{align}
|\rho_{\tau\mu} c_\gamma| \lesssim 0.1\lambda_\tau.
\end{align}
So $\rho_{\tau\mu} \sim 0.2$ gives $c_\gamma \lesssim 0.005$,
which is close to the alignment limit.
In this paper we take the alignment limit of 
$c_\gamma \to 0$ and $s_\gamma \to -1$ to simplify.

\begin{figure}[t!]
\center
\includegraphics[width=.44\textwidth]{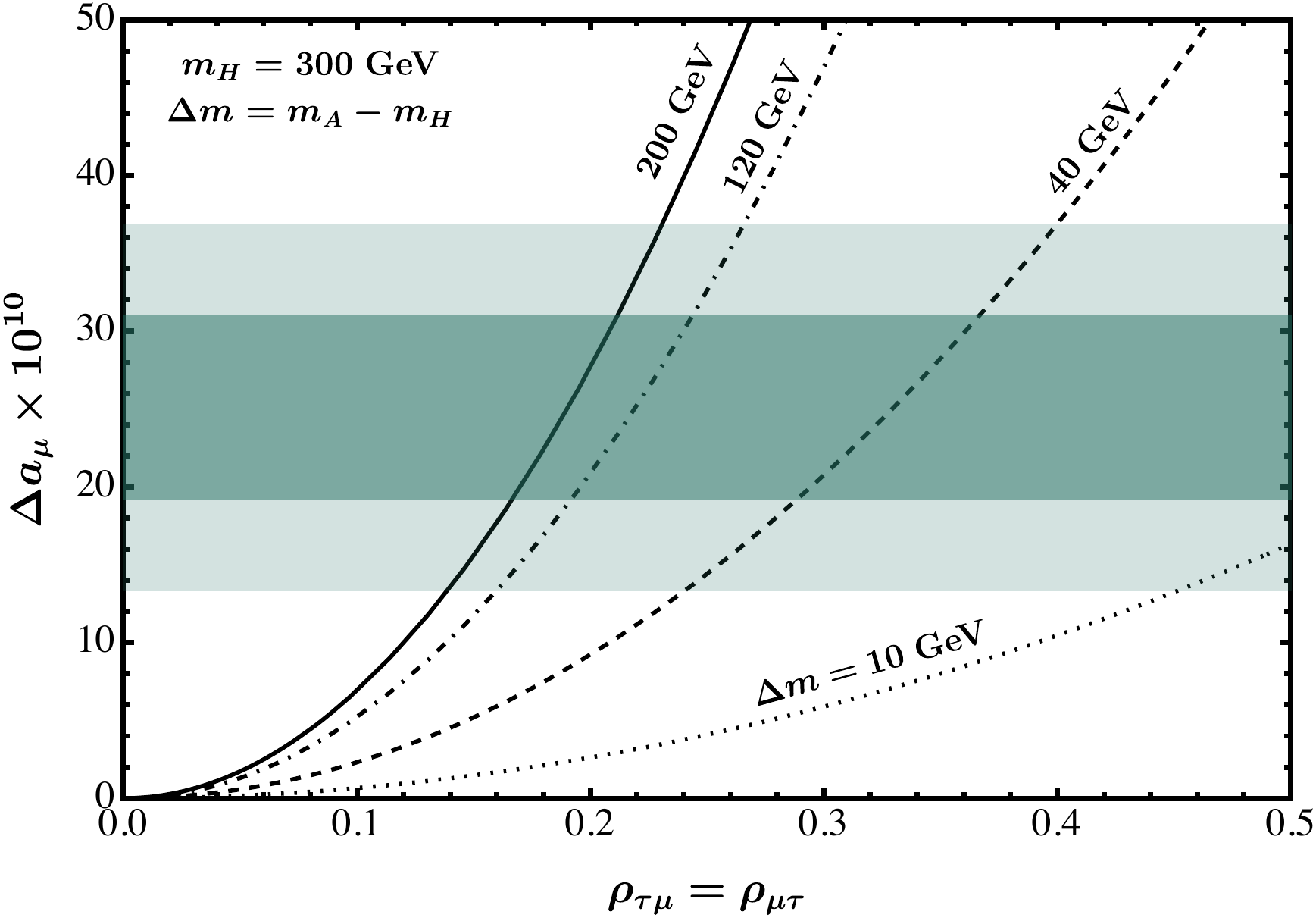}
\caption{
 $\Delta a_\mu$ vs $\rho_{\tau\mu} = \rho_{\mu\tau}$ for
 $m_H = 300$ GeV and $\Delta m = m_A - m_H = 10$, 40, 120, 200~GeV,
 with experimental $1\sigma$ and $2\sigma$ ranges (Eq.~(\ref{da_mu}))
 as indicated.
}
\label{Dela}
\end{figure}

Setting $c_\gamma = 0$ and using the well-known one-loop formula~\cite{Iguro:2019sly,Assamagan:2002kf,Davidson:2010xv,Omura:2015nja,Omura:2015xcg,Iguro:2017ysu},
we plot $\Delta a_\mu$ vs $\rho_{\tau\mu} = \rho_{\mu\tau}$ for $m_H = 300$~GeV
 and $\Delta m = m_A - m_H = 10$, 40, 120, 200~GeV in Fig.~\ref{Dela}, 
together with the 1$\sigma$ and $2\sigma$ ranges from Eq.~(\ref{da_mu}).
We note that $A$ and $H$ exactly cancel one another in $\Delta a_\mu$ 
when degenerate, which is illustrated by $\Delta m = 10$~GeV. 
Such near degenerate masses would require extremely large $\rho_{\tau\mu}$ values.
With $m_H = 300$~GeV as benchmark, the three values of 
$m_A = 340,\; 420,\;500$ GeV are taken for illustration.
One sees that, as $m_A$ increases toward decoupling, 
$H$ starts to dominate and the $\rho_{\tau\mu}$ strength needed
for 1$\sigma$ solution decreases, namely
\begin{align}
\rho_{\tau\mu} \gtrsim 0.289,\, 0.192,\, 0.167,
\label{rhotaumu}
\end{align}
which are $\sim 20$ times larger than $\lambda_\tau \simeq 0.01$.

If one replaces the muon in the left of Fig.~\ref{feyndiag}(left) by a tau,
a finite $\rho_{\tau\tau}$ can induce $\tau\to \mu\gamma$.
We have checked  that for $\rho_{\tau\tau}  = {\cal O}(\lambda_\tau)$
{or smaller~\cite{rho_tautau}},
the one-loop chiral suppression gives a rate that is quite below existing bounds~\cite{PDG}.
It is known~\cite{Davidson:2010xv} that a two-loop mechanism 
can in fact dominate for appropriate $\rho_{tt}$ values in the fermion loop,
as seen from Fig.~\ref{feyndiag}(right).
The mechanism is similar to $\mu \to e\gamma$~\cite{Chang:1993kw}, 
from which the formulas are adapted.
Using the known formulas and 1$\sigma$ solution values of
Eq.~(\ref{rhotaumu}), and assuming $\rho_{\tau\tau} \sim 0$,
one finds $|\rho_{tt}\rho_{\tau\mu}| < 0.014,\, 0.016,\, 0.017$
for $m_H = 300$~GeV and $m_A = 340,\, 420,\, 500$~GeV, or 
\begin{align}
|\rho_{tt}|  \lesssim  0.05,\, 0.08,\, 0.1,
\label{rhott_tamuga}
\end{align}
which are rather small compared with $\lambda_t \cong 1$.
With $\rho_{tt}$ considerably smaller than 1, it is generally easier
to accommodate flavor constraints,
especially in the quark sector~\cite{Altunkaynak:2015twa,Hou:2020itz},
and we do not explore it further.
%
%

As we will show in the next section, search for
$gg \to H,\, A \to \tau\mu$ at the LHC turns out to provide
more stringent bound on $\rho_{tt}$ than Eq.~(\ref{rhott_tamuga}).
%
One thought~\cite{Iguro:2019sly} then is to dilute ${\cal B}(H,\, A \to \tau\mu)$
with a second extra top Yukawa coupling, namely $\rho_{tc}$.
However, $B \to D\mu\nu$ vs $B \to De\nu$ universality
provides a bound~\cite{Iguro:2019sly} on $\rho_{tc}\rho_{\tau\mu}$ 
through $H^+$ exchange, where the $H^+\bar cb$ and $H^+\mu\bar\nu_\tau$ 
couplings can be read off Eq.~(\ref{eq:Yuk}). 
The not so intuitive point is that the former is 
$\simeq V_{tb}\rho_{tc}$~\cite{Ghosh:2019exx} and not CKM suppressed!
Thus, $\rho_{tc}\rho_{\tau\mu}$ would generate $B \to D\mu\bar\nu_\tau$,
which would add incoherently~\cite{Iguro:2017ysu} to $B \to D\mu\bar\nu$ 
that is measured by experiment rather precisely, but cannot discern the neutrino flavor. 
From Eq.~(\ref{rhotaumu}) and using Eq.~(84) of Ref.~\cite{Iguro:2017ysu},
we compare with the Belle measurement~\cite{Glattauer:2015teq},
\begin{align}
{\cal B}(B \to D \mu \nu)/{\cal B}(B \to D e \nu)
  = 0.995 \pm 0.022 \pm 0.039.
\label{BtoDlnu}
\end{align}
Adding errors in quadrature and allowing $2\sigma$ range,
we find $|\rho_{tc}\rho_{\tau\mu}| \lesssim 0.024,\, 0.037,\, 0.053$
for $m_H = 300$~GeV and $m_{H^+} = m_A = 340$, 420, 500~GeV,
giving
\begin{align}
|\rho_{tc}|  \lesssim  0.08,\, 0.19,\, 0.32.
\label{rhotc_Dlnu}
\end{align}
The decreasing value of $\rho_{\tau\mu}$ with increasing $m_A$, 
viz. Eq.~(\ref{rhotaumu}), implies the opposite for $\rho_{tc}$, 
which for $m_A \gtrsim 400$~GeV can surpass the $\rho_{\tau\mu}$ 
strength needed for 1$\sigma$ solution of muon $g-2$.

In fact, a good part of the dilution effect for ${\cal B}(H,\, A \to \tau\mu)$ is 
driven by the opening of $A \to HZ$ decay for $m_A \gtrsim 400$ GeV, which 
is one reason why we increased $\Delta m = m_A - m_H > m_Z$.
With $m_A = m_{H^+}$, this means $H^+ \to HW^+$ also opens up.
As we shall see in the next section, a sizable $\rho_{tc}$ 
can give rise to $cg \to bH^+$ production~\cite{Ghosh:2019exx}
(again, the $\bar cbH^+$ coupling is not CKM-suppressed),
as well as~\cite{Kohda:2017fkn} $cg \to tH, tA$,
 with $cg \to tA$ suppressed for higher $m_A$.
$H^+ \to HW^+$ decay with $H \to t\bar c,\, \tau\mu$ would
lead to 
additional signatures at the LHC.




Let us comment on a few other flavor concerns.
With $\rho_{\tau\mu}$ sizable, it can induce $\tau \to 3\mu$ decay 
at tree-level with $\rho_{\mu\mu} \neq 0$. 
Thus, the bound by Belle~\cite{Hayasaka:2010np},
${\cal B}(\tau \to 3\mu) \lesssim 2.1 \times 10^{-8}$, 
puts a constraint on $\rho_{\mu\mu}$.
Using formulas from Refs.~\cite{Hou:2020itz,Crivellin:2013wna},
we find $\rho_{\tau\mu}\rho_{\mu\mu} \lesssim (240$--$320)\lambda_\tau\lambda_\mu$ 
for $m_H = 300$~GeV and $m_A - m_H = 10$--200 GeV.
This implies $\rho_{\mu\mu} \lesssim 10\lambda_\mu < 0.01$,
which is still rather small.
One evades $h \to \mu\mu$ search in alignment limit, $c_\gamma = 0$, 
while one also evades the recent $H,\, A \to \mu\mu$ search~\cite{Sirunyan:2019tkw} 
by CMS, {after the multi-mode dilution effects discussed in 
the next section are taken into account.}


The coupling $\rho_{\mu\tau}$\,($\rho_{\tau\mu}$) enters 
the $H^+\bar\nu_\mu\tau$\;($H^+\bar\nu_\tau\mu$) coupling directly, 
and can generate $\tau \to \mu\nu_\mu\bar\nu_\tau$,
where the neutrino flavors are swapped compared with $W$ boson exchange,
which the experiment cannot distinguish.
Compared with $\tau\, (\mu) \to e\nu\bar\nu$, this constitutes another test of
lepton universality violation at the per mille level, 
which has been recorded by HFLAV~\cite{Amhis:2019ckw}.
We have checked that $\rho_{\tau\mu} = \rho_{\mu\tau} \lesssim 0.9$ 
is still allowed for our benchmark masses, 
which is more accommodating than muon $g-2$.

Finally, the $\rho_{\tau\mu} = \rho_{\mu\tau}$ coupling
affects $Z \to \tau\tau,\, \mu\mu$ decays, 
which have been precisely measured~\cite{PDG}, again at the per mille level.
We find the bound to be weaker than the bounds from $\tau$ decays discussed here.

\begin{figure*}[t]
\center
\includegraphics[width=1 \textwidth]{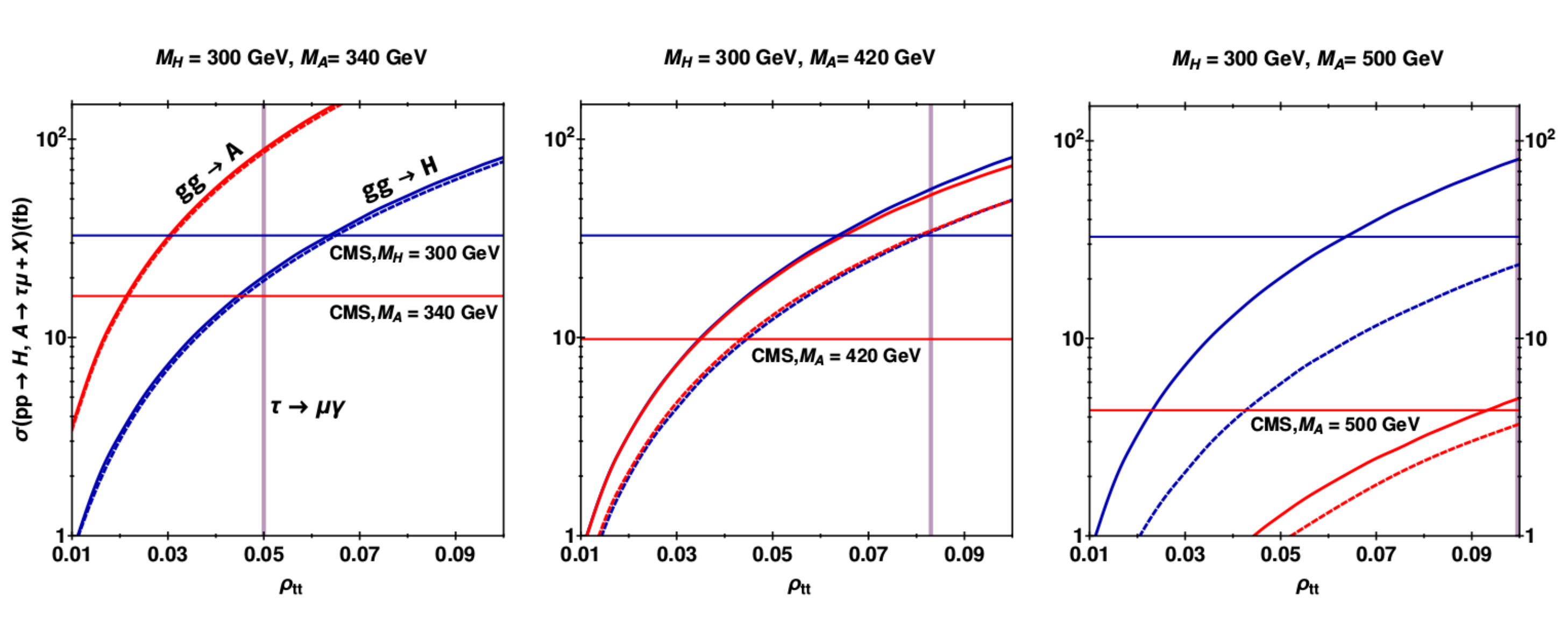}
\caption{
 Cross sections for $pp \to H\, (A) \to \tau\mu + X$ vs $\rho_{tt}$ at $\sqrt s = 13$~TeV,
 where blue (red) solid curves are for $\rho_{\tau\mu} \simeq $
 0.289 (left), 0.19 (center), 0.167 (right) from Eq.~(\ref{rhotaumu})
 with $\rho_{tc} = 0$ and $m_H = 300$ GeV, and
 for $m_A = 340$, 420, 500 GeV, respectively.
 The dashed curves correspond to $\rho_{tc} \simeq$ 0.08 (left), 0.19 (center), 0.32 (right)
 from Eq.~(\ref{BtoDlnu}). All other $\rho_{ij}^f$s are set to zero.
 Bounds from $\tau \to \mu\gamma$~\cite{Abdesselam:2021cpu}
 and CMS search~\cite{Sirunyan:2019shc} for $gg \to H \to \tau\mu$
 (reinterpreted in case of $A$) are indicated.
}
\label{fig:rhott_CMS_Htaumu}
\end{figure*}

\section{Flavor Probes vs LHC Search}


The connection between muon $g-2$ (or $\tau\to \mu\gamma$) with $H,\, A \to \tau\mu$ search~\cite{Iguro:2019sly,Assamagan:2002kf,Davidson:2010xv,Omura:2015nja,Omura:2015xcg,Iguro:2017ysu}  
at hadron colliders is well  known.
In part because of the $\tau \to \mu\gamma$ constraint on $\rho_{tt}$,
Ref.~\cite{Iguro:2019sly} did not explore $gg \to H,\, A \to \tau\mu$
but emphasized the electroweak pair production of exotic scalar boson pairs instead.
However, we are interested in keeping $\rho_{tt}$ as 
CMS has recently performed a search for 
heavy $H \to \tau\mu$ (and $\tau e$) channel. 
Using $\sim 36$ fb$^{-1}$ data at 13 TeV and assuming narrow width, 
the bound~\cite{Sirunyan:2019shc} on $\sigma(gg \to H){\cal B}(H \to \tau\mu)$ 
ranges from 51.9 fb to 1.6 fb for $m_H$ ranging from 200 GeV to 900 GeV.
The $H$ notation is intentional: CMS followed Ref.~\cite{Buschmann:2016uzg}
and studied the scalar $H$ case, but not for pseudoscalar $A$.

It is of interest to check the current LHC bound on $\rho_{tt}$
within the large $\rho_{\tau\mu} = \rho_{\mu\tau}$ interpretation of muon $g-2$.
We evaluate the $pp \to H,\, A \to \tau\mu + X$ cross section for LHC at 13~TeV,
assuming $gg \to H,\, A$ as the dominant subprocess and
convolute with parton distribution functions.
Taking $\rho_{tt}$ as real for our benchmark $m_H$ and $m_A$ values,
we fold in the decay branching ratios ${\cal B}(H,\, A \to \tau\mu)$
and give our results in {Fig.~\ref{fig:rhott_CMS_Htaumu}} 
as {blue\;(red)} solid curves for $H$\;($A$), 
with the corresponding CMS bound given as {horizontal lines} 
for our three benchmark sets.
The bounds from $\tau \to \mu\gamma$, Eq.~(\ref{rhott_tamuga})
are given as (dark pink) vertical lines.
We have checked that $\Gamma_H \sim 1$ GeV, 
while $\Gamma_A < 2$ GeV for $m_A < 400$ GeV, 
and remains below 10 GeV for $m_A \sim 500$ GeV, 
hence satisfy the narrow width approximation.
We do not combine the two separate states.

\begin{figure*}[t]
\center
\includegraphics[width=.33\textwidth]{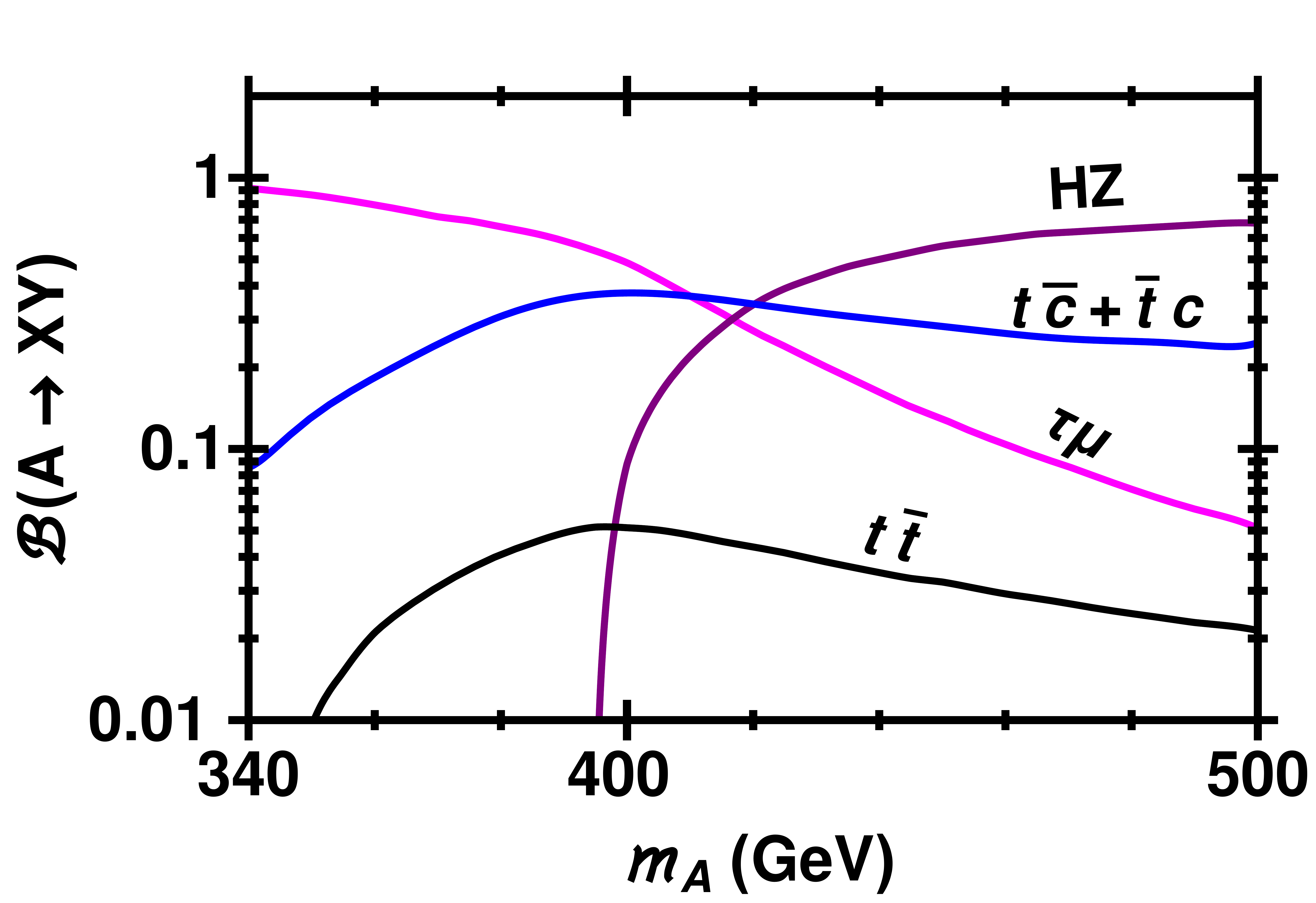}
\hskip0.15cm
\includegraphics[width=.33\textwidth]{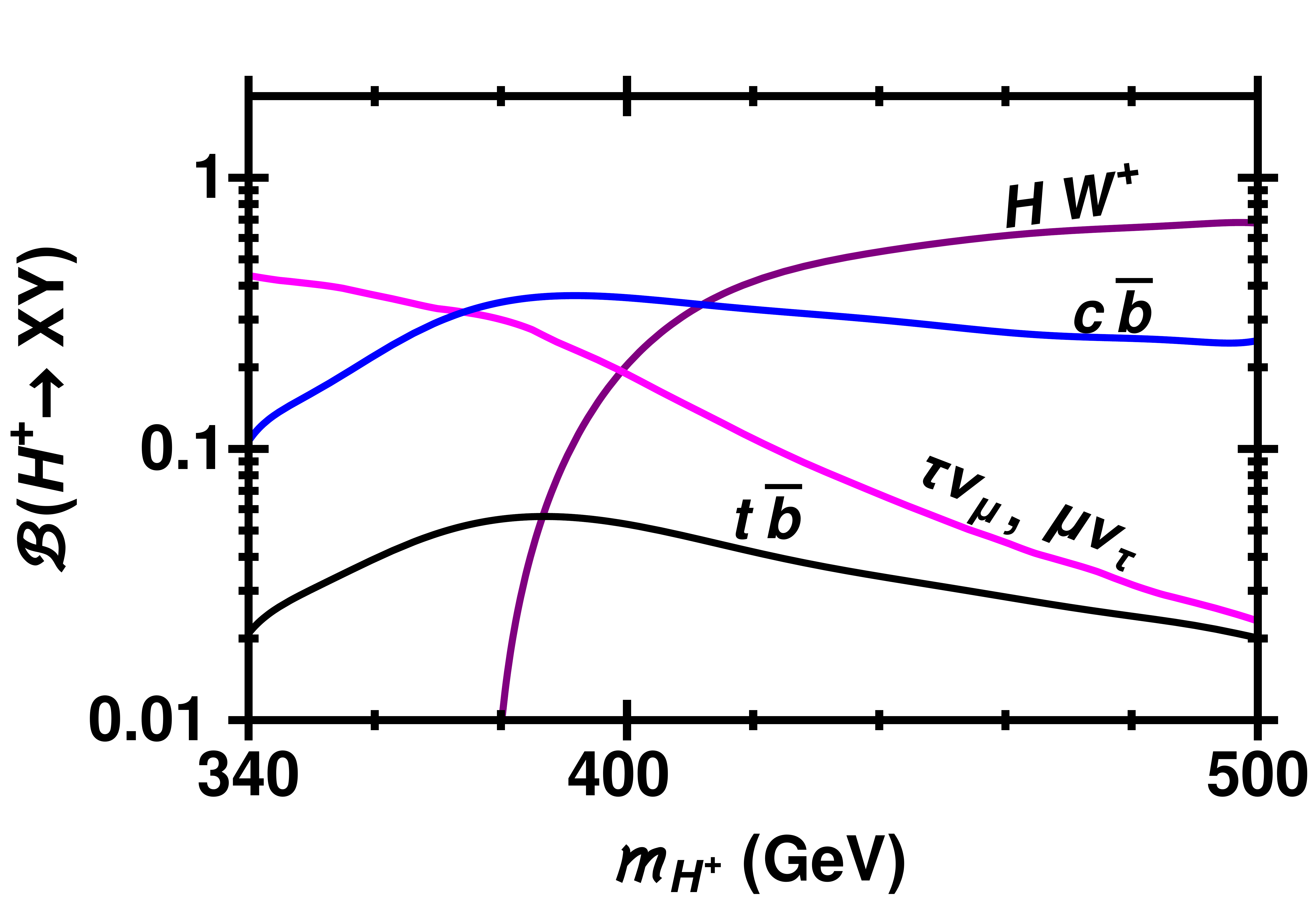}
\caption{
 Branching fractions for $A$ and $H^+$ decay modes vs mass from
 $\rho_{\tau\mu}$, $\rho_{tc}$, $\rho_{tt}$ (see Eq.~(\ref{eq:Yuk})) and
 the weak interaction, where $m_H = 300$ GeV and $m_A = m_{H^+}$ is assumed.
}
\label{fig:BR_AH+}
\end{figure*}

We see that, with only a subset of Run~2 data, the CMS bound
is more stringent than the bound from $\tau \to \mu\gamma$,
as one should always take the more stringent bound out of $H$ vs $A$.
For the cases of $m_A = 340$, 420 GeV, the larger production cross sections
for $A$ imply more stringent limit than from $H$,
but for $m_A = 500$ GeV, parton densities have dropped too low,
and the bound from $H$ is more stringent than $A$,
but still more stringent than from $\tau \to \mu\gamma$.
We remark that the case of $m_H,\, m_A = 300$, 340 GeV,
  {i.e. Fig.~\ref{fig:rhott_CMS_Htaumu}(left)}, 
does provide a $1\sigma$ solution to muon $g-2$.
But if all other extra Yukawa couplings, $\rho^f_{ij}$s ($f = u,\, d,\, \ell$),
are not stronger than $\lambda_{\rm max(i,\, j)}$, the Yukawa strength in SM,
then there are little other consequences,
except in the $gg \to A \to \tau\mu$ probe of $\rho_{tt}$ itself.
{The reinterpreted CMS bound of $\rho_{tt} \lesssim 0.02$ from Fig.~\ref{fig:rhott_CMS_Htaumu}(left) is below $\rho_{tt} \lesssim 0.05$
allowed by $\tau \to \mu\gamma$, 
which is really small compared with $\lambda_t \cong 1$.
{\it But there is still discovery potential with full Run~2 data}.
}
%
The allowed range for $\rho_{tt}$ from CMS gets partially restored
for heavier $m_A$, where $gg \to H,\, A \to \tau\mu$ production becomes
predominantly through $H$ in  {Fig.~\ref{fig:rhott_CMS_Htaumu}(right)}.
The easing of the bound on $\rho_{tt}$ is due to
the opening of $A \to t\bar t$, $HZ$,
as can be seen from the branching ratio plot,  {Fig.~\ref{fig:BR_AH+}(left)}.

In addition to $\rho_{tt}$ needed for $\tau \to \mu\gamma$ and $gg \to H,\, A$ production, 
we give the effect of turning on $\rho_{tc}$, 
the second extra top Yukawa coupling, in {Fig.~\ref{fig:BR_AH+}(left)},
as it can dilute the $gg \to H,\, A \to \tau\mu$ cross section.
However, as discussed in the previous section, the product of 
$\rho_{tc}\rho_{\tau\mu}$ can make $B \to D\mu\nu$ deviate 
from $B \to D e\mu$ and break $e$--$\mu$ universality.
From the 1$\sigma$ solution to muon $g-2$, Eq.~(\ref{rhotaumu}),
for our benchmark $m_H = 300$~GeV, $m_A = 340$, 420, 500~GeV, 
Belle data~\cite{Glattauer:2015teq} therefore leads to 
the bound on $\rho_{tc}$, giving Eq.~(\ref{BtoDlnu}).
Taking $\rho_{tc} \simeq 0.08,\, 0.19,\, 0.32$,
and the $\rho_{tc}$-induced dilution for the three corresponding 
$m_A$ values in {Fig.~\ref{fig:BR_AH+}(left)}, we draw 
the dashed curves in {Fig.~\ref{fig:rhott_CMS_Htaumu}}. 
Not much is changed in {Fig.~\ref{fig:rhott_CMS_Htaumu}(left)}, but for
 {Fig.~\ref{fig:rhott_CMS_Htaumu}(center) and Fig.~\ref{fig:rhott_CMS_Htaumu}(right)},
the $gg \to H,\, A \to \tau\mu$ cross section weakens progressively.
It is less significant for $A$ production at $m_A = 500$~GeV in 
Fig.~\ref{fig:rhott_CMS_Htaumu}(right), because $A \to HZ$ is much stronger.
But the corresponding drop in $gg \to H \to \tau\mu$ is rather significant. 
This is because the $\rho_{tc} \simeq 0.32$ value is about twice as large as
the $\rho_{\tau\mu}$ value of $\simeq 0.167$.
{The} large $\rho_{tc}$ brings about interest, thereby possible constraints,
from production processes at the LHC.

Before discussing $\rho_{tc}$-induced processes, we remark that 
{Fig.~\ref{fig:BR_AH+}}, where we also give $H^+$ decay branching fractions,
is not just for our three benchmark values of $m_A = 340$, 420, 500~GeV.
The plot is made by the same approach:
for $m_H = 300$~GeV and a given $m_A$,
we find the 1$\sigma$ muon $g-2$ solution value for $\rho_{\tau\mu}$
analogous to Eq.~(\ref{rhotaumu}),
then find the allowed upper bounds for $\rho_{tt}$ and $\rho_{tc}$,
analogous to Eqs.~(\ref{rhott_tamuga}) and (\ref{rhotc_Dlnu}), respectively.
The plot therefore scans through $m_A$, with $m_H$ fixed at 300~GeV.


Having $\rho_{tc}$ alone with $\rho_{tt}$ small, it can generate~\cite{Kohda:2017fkn} 
$cg \to tH,\, tA \to tt\bar c$ or same-sign top plus jet final state, which 
can~\cite{Hou:2019gpn} {feed the $t\bar tW$ control region (CRW) 
of} $4t$ search by CMS (by different selection cuts, ATLAS is less stringent),
where {there} is now a full Run~2 data study~\cite{Sirunyan:2019wxt}.
More significant is a study~\cite{Hou:2021xiq} of $cg \to bH^+ \to bAW^+$ production 
followed by $A \to t\bar c$ and both the top and the $W^+$ decay (semi-)leptonically,
which can also feed the CRW of CMS $4t$ search.
In this study, $H^+ \to AW^+$ decay was treated 
with $m_{H^+} = m_H$,
i.e. swapping $H \leftrightarrow A$ from our present case.
The $bH^+$ cross section is sizable due to a light $b$ quark and 
no CKM suppression for the $\bar c bH^+$ coupling~\cite{Ghosh:2019exx}.
It was found that $\rho_{tc} \sim 0.15$ is barely allowed for $m_{H^+} \sim 500$ GeV.
This may seem to make the large values of $\rho_{tc}$ for our dilution effect untenable.

However, the dilution effect may solve itself.
We note that, assuming just two associated extra Yukawa couplings, 
i.e. $\rho_{\tau\mu}$ and $\rho_{tc}$, the two branching ratios are
${\cal B}(H \to t\bar c + \bar t c) : {\cal B}(H \to \tau\mu) =
5\% : 95\%$, $40\% : 60\%$, $71\% : 29\%$, for 
$m_A = 340$, 420, 500 GeV and $(\rho_{\tau\mu},\, \rho_{tc}) = (0.289,\, 0.08)$, 
 $(0.192,\, 0.19)$, $(0.167,\, 0.32)$, respectively.
For the $m_A = m_{H^+} = 340$ GeV benchmark,
$\rho_{tc}$ is too small to be of concern.
Likewise, for $m_A = m_{H^+} = 420$ GeV benchmark,
{from Fig.~4 of Ref.~\cite{Hou:2020chc},
which is for $m_H = 300$ GeV and $m_A = 350$~GeV,
 CRW of CMS $4t$ search would push $\rho_{tc}$ to below 0.4.
However, $m_A$ at the much heavier 420 GeV 
reduces $cg \to tA \to tt\bar c$ events that feed CRW,
so one is not yet sensitive to $\rho_{tc} \simeq 0.2$,}
while there is some dilution from $A \to HZ$.
For the more efficient $cg \to bH^+$ production,
we see from {Fig.~\ref{fig:BR_AH+}(right)} that 
 {$H^+ \to c\bar b$ at $\simeq 32\%$ is pure dilution, and
 even $\tau\nu_{(\mu)} + \mu\nu_{(\tau)}$ at $\simeq 22\%$ together
 with the $b$ would likely be overwhelmed by single-top background. 
 The dominant ($\simeq 41\%$) $bHW^+$ final state} would be
 20\% $bt\bar cW^+$ and 60\% $b\tau^\pm\mu^\mp W^+$.
 The latter would be an interesting signature in itself,
 but would not feed CRW of CMS $4t$ search.

It is the $m_A = m_{H^+} = 500$ GeV benchmark
 that CRW of CMS $4t$ may put a limit on $\rho_{tc}$:
 $H^+ \to c\bar b$ is slightly below $30\%$, while
 $\tau\nu_{(\mu)} + \mu\nu_{(\tau)}$ is no longer a dilution factor.
 This is due to the dominance of $H^+ \to HW^+$, now at 70\%.
 With $H \to t\bar c$ at $\sim 35\%$, it would feed $4t$ CRW abundantly,
 as pointed out in Ref.~\cite{Hou:2021xiq} for the analogous twisted custodial case
 for $bAW^+$ production. If we take the number there as a guesstimate,
 then $\rho_{tc} \sim 0.15$ would be borderline.
This just means that $\rho_{tc}$ should be considerably less than
0.32 from Eq.~(\ref{BtoDlnu}), which means that one cannot reach $\rho_{tt} \sim 0.1$
as allowed by $\tau \to \mu\gamma$. 

We note that, from Fig.~\ref{mHmA_scan} and for $m_H \sim 300$ GeV,
$m_A = m_{H^+}$ at 500 GeV is already at the ``border'' of the scan space.
Thus, we conclude that for large splitting of $m_{A} = m_{H^+}$ from $m_H$, 
the solution to muon $g-2$ would not allow a $\rho_{tc}$
value larger than $\rho_{\tau\mu}$ for the one-loop solution.
As one moves $m_H$ higher, the discussion should be similar
for higher $m_A = m_{H^+}$, requiring larger $\rho_{\tau\mu}$ values
to solve muon $g-2$.



\section{Discussion and Conclusion}

Referring to the $4.2\sigma$ ``white space of disagreement'' between 
theory and experiment, the question ``What monsters may be lurking there?'' 
from the April 7 announcement presentation by the Muon $g-2$ experiment, 
became the quote of the day.
We stress that $\rho_{\tau\mu}/\lambda_\tau \sim 20$, while large,
is not ``monstrous'', and is up to {\it Nature} to choose
 {(and the counterbalancing $\rho_{tt}/\lambda_t \lesssim 1/10$ as well)}.
This can be compared, for example, with $\tan\beta = {\cal O}(10^3)$
in the muon-specific 2HDM~\cite{Abe:2017jqo}, tailor-made 
for the muon $g-2$ by some $Z_4$ symmetry.
In a different $Z_4$ arrangement, the $\mu$-$\tau$-philic 
extra scalar doublet~\cite{Abe:2019bkf} was introduced 
with only the $\rho_{\tau\mu} = \rho_{\mu\tau}$ coupling,
with further variations of~\cite{Wang:2019ngf} $m_A - m_H < 50$~GeV 
by considering the $\sim 2\sigma$ deviation from SM in $\tau$ decays,
or pushing $m_H$ to be very light~\cite{Wang:2021fkn}.
We hope we have illustrated that such elaborations are not necessary,
that the g2HDM is versatile enough with its arsenal of extra Yukawa couplings, 
to provide solution to muon $g-2$ while hiding from view so far,
but with promising LHC implications.

With enhanced $\rho_{\tau\mu}$ implied by muon $g-2$,
it has been suggested~\cite{Chiang:2016vgf,Guo:2016ixx} that, 
by the natural complexity of extra Yukawa couplings,  
they could possibly drive electroweak baryogenesis (EWBG),
i.e. be the $CP$ violating source of the baryon asymmetry of the Universe.
In fact, part of our motivation for considering 
$\rho_{tt}$ and $\rho_{tc}$ are their potential
 as drivers~\cite{Fuyuto:2017ewj} for EWBG.
Our findings show that $\rho_{tt}$ may still suffice,
but $\rho_{tc}$ may not be strong enough.
But whether one deals with large $\rho_{\tau\mu}$
or small but finite $\rho_{tt}$ as the EWBG driver, 
one may have to face the issue of the alignment limit, 
or extremely small $h$--$H$ mixing.
{
We note that, by the arguments we have given,
the simultaneously large $\rho_{\tau\mu}$, $\rho_{\tau\tau}$ 
and $\rho_{tt}$ as given in Ref.~\cite{Chiang:2016vgf}
cannot survive collider bounds such as from
CMS search for $H,\, A \to \tau\mu$~\cite{Sirunyan:2019shc},
and ATLAS~\cite{ATLAS:2020zms} and CMS~\cite{Sirunyan:2018zut} 
searches for $H,\, A \to \tau\tau$, which we have checked explicitly.
}

We note in passing that g2HDM has been invoked in weaker
 (considered as a perturbation hence keeping $\tan\beta$) form~\cite{Ghosh:2020tfq}, 
where having rather small $\rho_{\tau\mu}$ called for relatively light $m_A$.
In a similar vein, a reinterpretation~\cite{Primulando:2019ydt} of 
the CMS $gg \to H \to \tau\mu$ search~\cite{Sirunyan:2019shc} 
also urged for probing below $m_H = 200$ GeV.
These suggestions are not invalid, but we chose $m_H$ within 
the search range of CMS at a typical weak scale 
to illustrate how g2HDM may resolve muon $g-2$.

The flavor conserving coupling $\rho_{\mu\mu}$ can in principle induce 
new contributions to muon $g - 2$ via one- and two-loop diagrams. 
The one-loop effect~\cite{Dedes:2001nx} involves $\rho_{\mu\mu}$ only, 
but because of chiral suppression from muon in the loop,
exorbitantly large $\rho_{\mu\mu} \gtrsim 0.85$ ($\sim 1400\lambda_\mu$: monstrous!) 
is needed for $1\sigma$ solution to muon $g-2$ for the same benchmark masses as before.
The two-loop effect~\cite{Barr:1990vd,Ilisie:2015tra}, 
by replacing the $\tau$ on the left of Fig.~\ref{feyndiag}(right) 
with $\mu$, can be enhanced by extra top Yukawa coupling $\rho_{tt}$.
But it again calls for the product of $\rho_{tt}\rho_{\mu\mu}$ to be too large.
Taking, for example, $\rho_{tt} \sim -0.1$   
for $m_H = 300$ GeV, $m_A = m_{H^+} = 340$ GeV, 
one can achieve 1$\sigma$ solution to muon $g-2$ for 
$\rho_{\mu\mu} \gtrsim 0.42$, which is again rather large~\cite{foot}. 
Indeed, this runs into conflict with the CMS search~\cite{Sirunyan:2019tkw} 
(the ATLAS bound~\cite{Aaboud:2019sgt} at similar data size is weaker)
for $gg \to H,\, A \to \mu^+\mu^-$ with $\sim 36$ fb$^{-1}$, 
which restricts  $\rho_{tt} \lesssim 10^{-2}$ for such a strong $\rho_{\mu\mu}$ value.
This shows, by power of the CMS bound~\cite{Sirunyan:2019tkw}
on $gg \to H,\, A \to \mu^+\mu^-$  even with just a fraction of LHC Run~2 data, 
that the two-loop $\rho_{\mu\mu}$ mechanism is not viable.

Finally, let us comment on the electroweak production processes
suggested in Ref.~\cite{Iguro:2019sly}, namely
$pp \to HA,\, H^\pm H,\, H^\pm A,\, H^+H^-$ through vector bosons.
For large $\rho_{\tau\mu}$ values needed for muon $g-2$, dileptonic decays 
of $H,\, A,\, H^\pm$ may prevail, which allow exquisite 
reconstruction~\cite{Iguro:2019sly} of exotic extra Higgs boson masses,
hence rather attractive.
However, the electroweak production cross section, 
at the fb level, is rather small.
Take the case of $m_H = 300$ GeV, $m_A = m_{H^+} = 340$ GeV for example.
If we allow the lowest $\rho_{tc}$ value of 0.08 in Eq.~(\ref{rhotc_Dlnu}),
nominally from $B \to D\ell\nu$ universality constraint,
we find that the $cg \to bH^+$ cross section,
being a strong production process involving a gluon
 (with not too significant a suppression from charm parton distributions) 
and only one heavy particle in final state, is almost two 
orders of magnitude higher than the electroweak production processes.
This effect persists for the other two benchmarks 
where {$A$ and $H^+$ are heavier}, which allow for larger $\rho_{tc}$ values. 
Furthermore, as we have illustrated, multiple decay modes of
the heavier $A$ and $H^+$ (see Fig.~\ref{fig:BR_AH+}), 
as well as $H \to \tau^\pm\mu^\mp,\, t\bar c\, (\bar tc)$
would dilute one another, which works also for the decays of 
the extra scalar bosons in electroweak pair production.
Thus, neither the search for the electroweak production of 
extra Higgs boson pairs, nor the reconstruction of the extra boson masses, 
may be as rosy as argued in Ref.~\cite{Iguro:2019sly}.

In summary, 
we employ large CLFV Yukawa coupling $\rho_{\tau\mu} = \rho_{\mu\tau} \sim 0.2$
in the general 2HDM to account for muon $g-2$ anomaly 
through the known one-loop mechanism.
The extra Higgs bosons have mass at the weak scale, but one is 
close to the alignment limit of very small $h$--$H$ mixing to evade $h \to \tau\mu$.
This motivates the check with $gg \to H,\,A \to \tau\mu$ search,
where a subset of LHC Run~2 data already puts more stringent bound 
on the extra top Yukawa coupling $\rho_{tt}$ than from
$\tau \to \mu\gamma$ through the two-loop mechanism.
The stringent constraint on $\rho_{tt}$ can be eased by
allowing a second extra top Yukawa coupling, $\rho_{tc}$,
which motivates the search for $cg \to bH^+ \to \tau^\pm\mu^\mp bW^+$,
$t\bar c bW^+$ (same-sign dilepton plus two $b$-jets and additional jet,
with missing $p_T$) at the LHC, on top of 
possible same-sign top plus jet signatures from $cg \to tH, tA \to tt\bar c$.
%
%
{Whether these extra Yukawa couplings can drive electroweak baryogenesis
should be further studied.
}

\vskip0.2cm
\noindent{\bf Acknowledgments} \
The work of WSH is supported by MOST 109-2112-M-002-015-MY3 of Taiwan 
and NTU 110L104019 and 110L892101, and
the work of RJ and GK are supported by {MOST 109-2811-M-002-565 
 and 109-2811-M-002-540, respectively.}
The {research of CK was supported in part by 
the U.S. Department of Energy and the University of Oklahoma}, and 
TM by a postdoctoral research fellowship from the Alexander von Humboldt Foundation.



\begin{thebibliography}{99}



%
\bibitem{Abi:2021gix}
  B.~Abi \textit{et al.} [Muon g-2],
  Phys. Rev. Lett. \textbf{126}, 141801 (2021).
%
\bibitem{Bennett:2006fi} 
  G.W.~Bennett {\it et al.} [Muon g-2],
  Phys.\ Rev.\ D {\bf 73}, 072003 (2006).
%
\bibitem{Aoyama:2020ynm} 
  T.~Aoyama {\it et al.},
  Phys.\ Rept.\  {\bf 887}, 1 (2020).
%
\bibitem{Borsanyi:2020mff} 
  Sz.~Borsanyi {\it et al.},
  Nature (2021).
%
\bibitem{Athron:2021iuf}
  P.~Athron, C.~Bal\'azs, D.H.~Jacob, W.~Kotlarski, D.~St\"ockinger
  and H.~St\"ockinger-Kim,
  arXiv:2104.03691 [hep-ph].
%
\bibitem{Iguro:2019sly}
  S.~Iguro, Y.~Omura and M.~Takeuchi,
  JHEP \textbf{11}, 130 (2019).
%
\bibitem{Assamagan:2002kf} 
  K.A.~Assamagan, A.~Deandrea and P.A.~Delsart,
  Phys.\ Rev.\ D {\bf 67}, 035001 (2003).
%
\bibitem{Davidson:2010xv} 
  S.~Davidson and G.J.~Grenier,
  Phys.\ Rev.\ D {\bf 81}, 095016 (2010).
%
\bibitem{Omura:2015nja} 
  Y.~Omura, E.~Senaha and K.~Tobe,
  JHEP {\bf 1505}, 028 (2015).
%
\bibitem{Omura:2015xcg} 
  Y.~Omura, E.~Senaha and K.~Tobe,
  Phys.\ Rev.\ D {\bf 94}, 055019 (2016).
%
\bibitem{Iguro:2017ysu} 
  S.~Iguro and K.~Tobe,
  Nucl.\ Phys.\ B {\bf 925}, 560 (2017).
%
%

\bibitem{Crivellin:2019dun}
A.~Crivellin, D.~M\"uller and C.~Wiegand,
JHEP \textbf{06}, 119 (2019).

\bibitem{Hou:2020chc} 
  For a brief review, see W.-S.~Hou and T.~Modak,
  Mod.\ Phys.\ Lett.\ A {\bf 36}, 2130006 (2021).
%
\bibitem{Glashow:1976nt} 
  S.L.~Glashow and S.~Weinberg,
  Phys.\ Rev.\ D {\bf 15}, 1958 (1977).
%
\bibitem{Hou:1991un} 
  W.-S.~Hou,
  Phys.\ Lett.\ B {\bf 296}, 179 (1992).
%
\bibitem{Cheng:1987rs} 
  T.-P.~Cheng and M.~Sher,
  Phys.\ Rev.\ D {\bf 35}, 3484 (1987).
%
\bibitem{Khachatryan:2016vau} 
  G.~Aad {\it et al.} [ATLAS and CMS],
  JHEP {\bf 1608}, 045 (2016).
%
\bibitem{Hou:2017hiw} 
  W.-S. Hou, M.~Kikuchi,
  EPL {\bf 123}, 11001 (2018).
%
\bibitem{Khachatryan:2015kon} 
  V.~Khachatryan {\it et al.} [CMS],
  Phys.\ Lett.\ B {\bf 749}, 337 (2015).
%
\bibitem{PDG}
  P.A.~Zyla {\it et al.} [Particle Data Group],
  PTEP {\bf 2020}, 083C01 (2020).
%
\bibitem{Sirunyan:2021ovv} 
  A.M.~Sirunyan {\it et al.} [CMS],
  arXiv:2105.03007 [hep-ex].
%
\bibitem{Chen:2013qta} 
  K.-F.~Chen, W.-S.~Hou, C.~Kao and M.~Kohda,
  Phys.\ Lett.\ B {\bf 725}, 378 (2013).
%
\bibitem{Hou:2019grj} 
  W.-S.~Hou, R.~Jain, C.~Kao, M.~Kohda, B.~McCoy and A.~Soni,
  Phys.\ Lett.\ B {\bf 795}, 371 (2019).
%
\bibitem{Abdesselam:2021cpu} 
  A.~Abdesselam {\it et al.} [Belle],
  arXiv:2103.12994 [hep-ex].
%
\bibitem{Sirunyan:2019shc} 
  A.M.~Sirunyan {\it et al.} [CMS],
  JHEP \textbf{03}, 103 (2020).
%
\bibitem{Davidson:2005cw} 
  S.~Davidson and H.E.~Haber,
  Phys.\ Rev.\ D {\bf 72}, 035004 (2005).
%
\bibitem{Hou:2019mve} 
  W.-S.~Hou and T.~Modak,
  Phys.\ Rev.\ D {\bf 101}, 035007 (2020).
%
\bibitem{Hou:2021xiq} 
  W.-S.~Hou and T.~Modak,
  Phys.\ Rev.\ D {\bf 103}, 075015 (2021).
%
\bibitem{Eriksson:2010zzb} 
  D.~Eriksson, J.~Rathsman and O.~St\aa l,
  Comput.\ Phys.\ Commun.\  {\bf 181}, 833 (2010).
%
{\bibitem{rho_tautau}
  Having checked this, we set $\rho_{\tau\tau}$ to zero
  in our numerical analysis for simplicity.
} 
%
\bibitem{Chang:1993kw} 
  D.~Chang, W.-S.~Hou and W.-Y.~Keung,
  Phys.\ Rev.\ D {\bf 48}, 217 (1993).
%
\bibitem{Altunkaynak:2015twa}
B.~Altunkaynak, W.~S.~Hou, C.~Kao, M.~Kohda and B.~McCoy,
Phys. Lett. B \textbf{751}, 135-142 (2015).
%
\bibitem{Hou:2020itz} 
  W.-S.~Hou and G.~Kumar,
  Phys.\ Rev.\ D {\bf 102}, 115017 (2020).
%
\bibitem{Ghosh:2019exx} 
  D.K.~Ghosh, W.-S.~Hou and T.~Modak,
  Phys.\ Rev.\ Lett.\  {\bf 125}, 221801 (2020).
%
\bibitem{Glattauer:2015teq} 
  R.~Glattauer {\it et al.} [Belle],
  Phys.\ Rev.\ D {\bf 93}, 032006 (2016).
%
\bibitem{Kohda:2017fkn} 
  M.~Kohda, T.~Modak and W.-S.~Hou,
  Phys.\ Lett.\ B {\bf 776}, 379 (2018).
%
\bibitem{Hayasaka:2010np}
  K.~Hayasaka, K.~Inami, Y.~Miyazaki \textit{et al.}
  Phys. Lett. B \textbf{687}, 139 (2010).
%
\bibitem{Crivellin:2013wna}
  A.~Crivellin, A.~Kokulu and C.~Greub,
  Phys. Rev. D \textbf{87}, 094031 (2013).
%
\bibitem{Sirunyan:2019tkw} 
  A.M.~Sirunyan {\it et al.} [CMS],
  Phys.\ Lett.\ B {\bf 798}, 134992 (2019).
%
\bibitem{Amhis:2019ckw}
  Y.S.~Amhis \textit{et al.} [HFLAV],
  Eur. Phys. J. C \textbf{81}, 226 (2021).
%
\bibitem{Buschmann:2016uzg}
  M.~Buschmann, J.~Kopp, J.~Liu and X.-P.~Wang,
  JHEP \textbf{06}, 149 (2016).
%
\bibitem{Hou:2019gpn} 
  W.-S.~Hou, M.~Kohda and T.~Modak,
  Phys.\ Lett.\ B {\bf 798}, 134953 (2019).
%
\bibitem{Sirunyan:2019wxt}
  A.M.~Sirunyan \textit{et al.} [CMS],
  Eur. Phys. J. C \textbf{80}, 75 (2020).
%
%
\bibitem{Abe:2017jqo}
  T.~Abe, R.~Sato and K.~Yagyu,
  JHEP \textbf{07}, 012 (2017).
%
\bibitem{Abe:2019bkf}
  Y.~Abe, T.~Toma and K.~Tsumura,
  JHEP \textbf{06}, 142 (2019).
%
\bibitem{Wang:2019ngf}
  L.~Wang and Y.~Zhang,
  Phys. Rev. D \textbf{100}, 095005 (2019).
%
\bibitem{Wang:2021fkn}
  H.-X.~Wang, L.~Wang and Y.~Zhang,
  [arXiv:2104.03242 [hep-ph]].
%
\bibitem{Chiang:2016vgf} 
  C.~W.~Chiang, K.~Fuyuto and E.~Senaha,
  Phys.\ Lett.\ B {\bf 762}, 315 (2016).
%
\bibitem{Guo:2016ixx}
  H.-K.~Guo, Y.-Y.~Li, T.~Liu, M.~Ramsey-Musolf and J.~Shu,
  Phys. Rev. D \textbf{96}, 115034 (2017).
%
\bibitem{Fuyuto:2017ewj} 
  K.~Fuyuto, W.-S.~Hou and E.~Senaha,
  Phys.\ Lett.\ B {\bf 776}, 402 (2018).
%
\bibitem{ATLAS:2020zms}
G.~Aad \textit{et al.} [ATLAS],
Phys. Rev. Lett. \textbf{125}, no.5, 051801 (2020)
%
\bibitem{Sirunyan:2018zut}
  A.M.~Sirunyan \textit{et al.} [CMS],
  JHEP \textbf{09}, 007 (2018).
%
\bibitem{Ghosh:2020tfq} 
  N.~Ghosh and J.~Lahiri,
  Phys.\ Rev.\ D {\bf 103},  055009 (2021).
%
\bibitem{Primulando:2019ydt} 
  R.~Primulando, J.~Julio and P.~Uttayarat,
  Phys.\ Rev.\ D {\bf 101}, 055021 (2020).
%
\bibitem{Dedes:2001nx}
  See e.g. A.~Dedes and H.E.~Haber,
   JHEP \textbf{05}, 006 (2001).
%
\bibitem{Barr:1990vd}
  S.M.~Barr and A.~Zee,
  Phys. Rev. Lett. \textbf{65}, 21 (1990).
%
\bibitem{Ilisie:2015tra}
  V.~Ilisie,
  JHEP \textbf{04}, 077 (2015); and references therein.
%
\bibitem{foot}
  The sign of $\rho_{tt}\rho_{\mu\mu}$ needs to be negative for
  generating muon $g-2$ by a pure two-loop mechanism.
  But the rather large $\rho_{\mu\mu}$ would have its own
  one-loop effect, where one can play with cancellations. 
  We do not get into this, but deem a purely two-loop mechanism
  as not viable.
%
\bibitem{Aaboud:2019sgt}
  M.~Aaboud \textit{et al.} [ATLAS],
  JHEP \textbf{07}, 117 (2019).


\end{thebibliography}
\end{document}